\documentclass[twocolumn,showpacs,preprintnumbers,amsmath,amssymb]{revtex4}

\usepackage{graphicx}
\usepackage{dcolumn}
\usepackage{bm}

\begin{document}

\preprint{APS/123-QED}

\title{A Multi-Chamber System for Analyzing the Outgassing, Deposition,\\and Associated Optical Degradation Properties of Materials in a Vacuum}

\author{Jack Singal}
 \email{jsingal@stanford.edu}
\author{Rafe Schindler}
\author{Chihway Chang}
\author{Patrick Czodrowski}
\author{Peter Kim}
\affiliation{Kavli Institute for Particle Astrophysics and Cosmology \\ SLAC National Accelerator Laboratory and Stanford University\\ Menlo Park, CA 94025}

\date{\today}

\begin{abstract}
We report on the Camera Materials Test Chamber, a multi-vessel apparatus which analyzes the outgassing consequences of candidate materials for use in the vacuum cryostat of a new telescope camera.  The system measures the outgassing products and rates of samples of materials at different temperatures, and collects films of outgassing products to measure the effects on light transmission in six optical bands.  The design of the apparatus minimizes potential measurement errors introduced by background contamination.
\end{abstract}

\maketitle

\section{\label{Intro}Introduction}

The Large Synoptic Survey Telescope (LSST) is a planned facility which will repeatedly and deeply image large (10 square degree) sections of the night sky at optical and very near infrared wavelengths to an unprecidented depth, gathering important data to address outstanding questions in cosmolgy and astrophysics\cite{LSSTover}.  The etendue, or light gathering ability, of LSST, as measured by the primary mirror aperture size multiplied by the field of view, is more than an order of magnitude higher than any existing or planned optical facility.  Such imaging requires a 3.2 billion pixel cryogenic CCD camera maintained in a clean vacuum environment.  

The cryostat of the LSST camera will measure 2.9 $m^{3}$, contain nearly 1000 kg of material, and have more than 40 $m^{2}$ of exposed surface area within, including the cold focal plane CCD detectors, two stages of supporting electronics, mechanical and thermal control structures, electrical cabling, electrical and fluid connectors and feed-thrus, and devices for internal metrology.  All materials within the evacuated cryostat will remain there for years and be subject to thermal cycling.  These materials must not outgass in a way that compromises the insulating vacuum nor change the net light transmittance to the focal plane between calibrations.  

There is scant data in the literature in regard to how the outgassing and deposition of commercial and other materials may affect light transmittance to optical surfaces in vacuum.  Abromovici et al. limited the effect of elastomer outgassing on mirror reflectivities\cite{ocon}.  They use resonant cavities pumped by a laser, thus measuring at only one fequency, and do not cool the optical collecting surfaces (mirrors).  Others have measured the outgassing and deposition rates of various matrials for many applications.  However, given the almost endless variety of available materials, achieving a database relevant to very different applications is impossible.

We have designed, contstructed, and commissioned an apparatus, the Camera Materials Test Chamber (CMTC), to quantiy the suitability of candidate materials for inclusion in the LSST camera cryostat.  The CMTC could also be used to perform suitability tests for any similar vacuum application, and, given its ability to isolate functions and cold stages, can be used for a variety of precision vacuum measurements, such as the performance of getters.

We report on the design and performance of the instrument, highlighting vacuum and thermal engineering considerations that may be of use to experimentalists.

\section{Experimental Setup} \label{setup}

\subsection{Setup overview} 

\begin{figure*}
\includegraphics[width=5.0in]{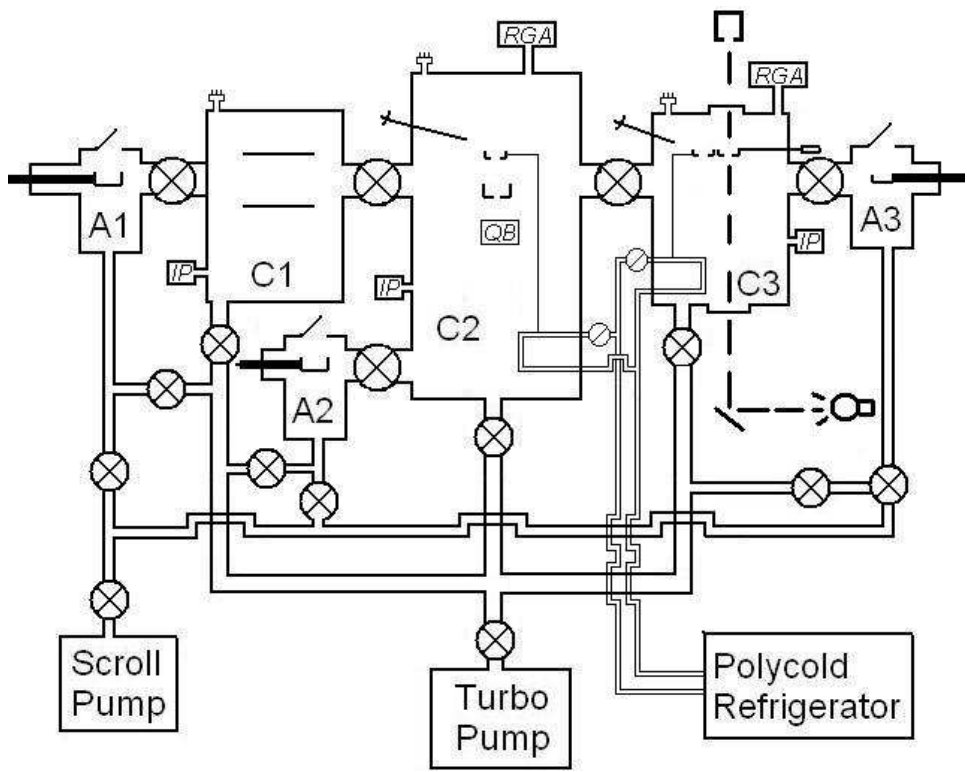}
\caption{Schematic of the CMTC, showing vessels, major devices, and vacuum and cryogen setup.  Abbreviations in the figure include "IP" for ion pump, "QB" for quartz microbalance mass analyzer, and "RGA" for residual gas analyzer.  Samples are heated in C1, and moved to C2 to outgas.  The rates for outgassing products are measured there with an RGA.  Depositions can be weighed with the quartz microbalance, and allowed to deposit onto a cold glass disk.  A contaminated glass disk is moved to C3 where the light transmittance through it is compared to an uncontaminated disk in six different wavelength bands.  Glass disks are maintained at -120$^{\circ}$ C in C2 and C3 via a thermal link to tubes containing cold fluid circulated by the refrigerator.  Samples enter through load lock A1 and exit through A2, while glass disks enter and exit through A3.  The RGA in C3 can be used to identify the condensed species on the disks. }
\label{schematic}
\end{figure*}

\begin{figure*}
\includegraphics[width=5.0in]{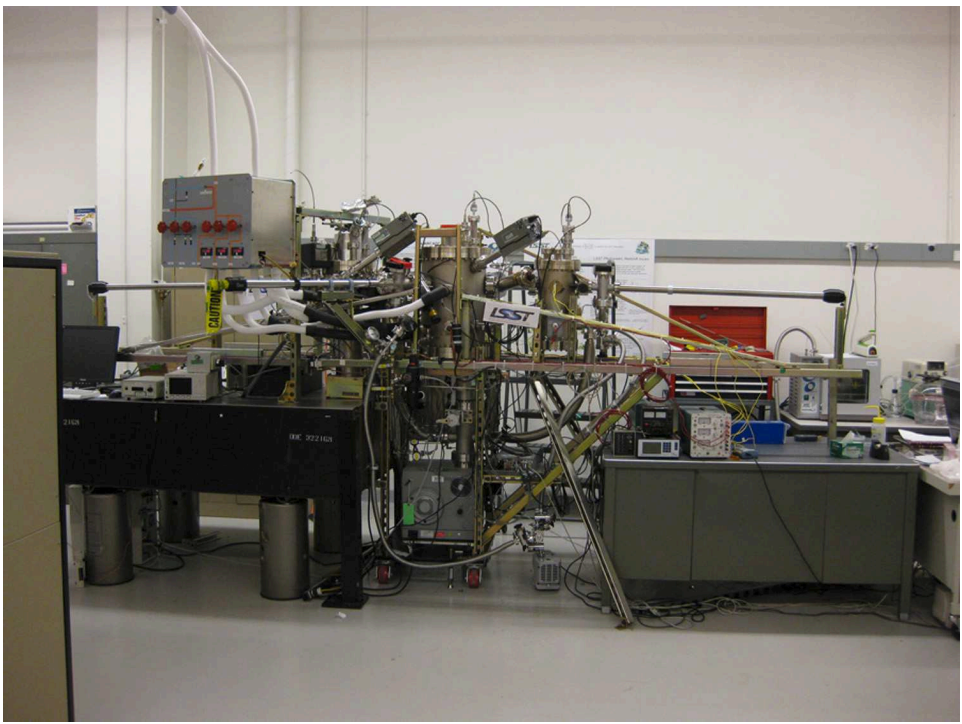}
\caption{Photograph of the entire CMTC system, showing vessels and supporting devices, and the physical scale.  The total setup is approximately 6 m long and 2 m high.  The magnetic transport arms are seen extending far to each side, along with supporting peripherals, including the optical table on the left.  The main vessels are in the middle.  The cryogenic valve box is the prominent structure mounted in the upper left, while part of the Polycold refrigeration unit is seen in the far left of the image.  }
\label{bigphoto}
\end{figure*}

In order to isolate different functions and achieve high vacuum when needed and low levels of contamination, the CMTC consists of three main vacuum vessels and three smaller ones.  Figure \ref{schematic} shows a schematic.  Small vessels A1, A2, and A3 serve as load locks for the main vessels C1, C2, and C3.  Three magnetic transporter arms provide movement between 1) A1, C1, and C2, 2) A2 and C2, and 3) A3, C3, and C2.  Figure \ref{bigphoto} shows a photograph of the entire CMTC apparatus.

Straight through or 'gate' valves are present at every junction between vessels, and all vessels can be independently  isolated from the others and the pumping system.  Wobble sticks with pincers in C2 and C3 allow the transfer of items between basket or platform structures on the ends of the long magnetic transporter arms and structures in the chambers, and a pneumatic actuator provides for repeated movement of a stage in C3.

The entire setup measures approximately 6 m long, and 2 m tall.  Much of the length is taken up by the magnetic transporter arms and gate valves between vessels.  The C2 vessel is 42 cm in diameter and 61 cm tall, while the C1 and C3 vessels are 25 and 30 cm in diameter and 48 and 30 cm tall, respectively.

\subsection{Procedure and description of vessels} \label{pdv}

\begin{figure}
\includegraphics[width=3.5in]{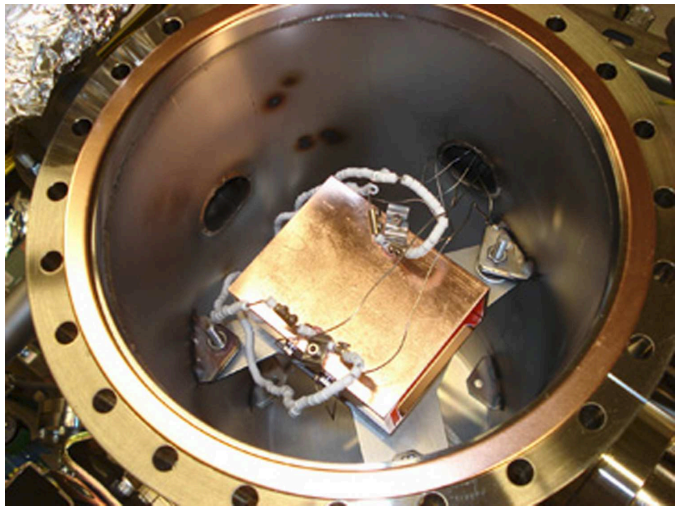}
\caption{Photograph of the interior of C1, showing the heater block and heater and thermocouple wiring. }
\label{C1photo}
\end{figure}

Small material samples (maximum 2.5 x 1 x 0.5 cm) are introduced into A1, where they are placed in a copper basket attached to the end of magnetic transport arm 1.  A1 is evacuated and the arm is then extended into C1, where the samples in the basket can be radiatively heated in an OFHC copper heater block under vacuum in accordance with how the material would be treated prior to integration into the camera.  The heater block is a 'tunnel' of OFHC copper of rectangular cross section into which the basket is extended.  It contains two hermetic heaters and two thermocouples.  Figure \ref{C1photo} shows a photograph of the interior of C1.  

\begin{figure}
\includegraphics[width=3.5in]{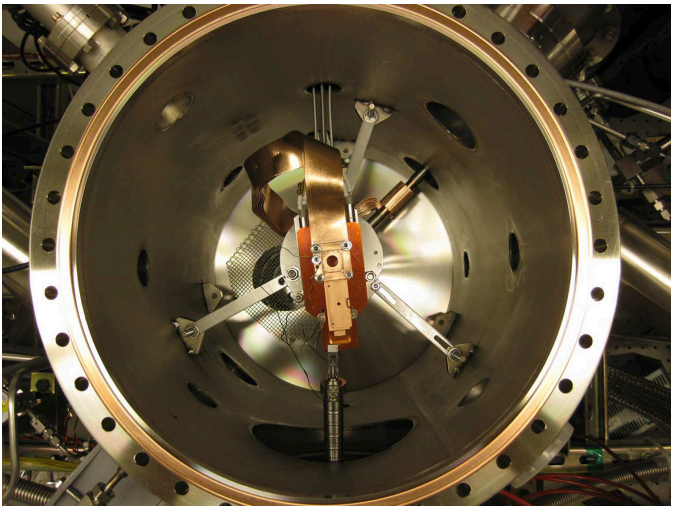}
\caption{Photograph of the interior of C2.  The glass disk stage is seen with thermal straps extending below to the cryogen loop.  The platform where the sample box slides is below the glass disk stage, and the box is in the install/remove position.  The cooling and data lines for the quartz microbalance scale disappear below the sample box platform.  The wobble stick extends into the frame from the bottom. }
\label{C2photo}
\end{figure}

Arm 1 is then further extended into C2 where a wobble stick is used to remove the material sample from the basket and place it in a small OFHC copper box.  The box has a sliding lid and small 2.5 mm diameter holes on the bottom and lid to direct outgassing flow.  Boxes enter and exit C2 through A2 via magnetic transport arm 2.  Figure \ref{C2photo} shows a photograph of the interior of C2.  

As seen in Figure \ref{C2photo}, the C2 sample box slides in a track in a copper platform supported from another platform on horizontal legs extending from bosses welded to the size walls of the chamber.   The box can either be positioned so that its lid can be open and samples placed within, or positioned so that the outgassing is directed through the holes to the glass disk stage above and a quartz microbalance deposition monitor mounted below.  The glass disk stage is mounted above the box platform with four 4-40 stainless steel allthread standoffs, and is connected via copper thermal straps to the cryogen loop near the bottom of the vessel.  The glass disk stage is maintained at -120$^{\circ}$C .  The quartz microbalance deposition monitor is supported by stainless steel tubes from its feed-thru flange and can also be cooled to -120$^{\circ}$C via a cryogen loop.  C2 contains thermocouples for reading temperatures, including on the box platform and on the cold glass disk stage.  

While in C2, the outgassing rate of all products by molecular weight up to 200 AMU is measured with a SRS RGA200 residual gas analyzer (RGA) using the rate-of-rise technique (see \S \ref{rorsec}).  The deposition rate of outgassing products from a sample with high outgassing can be measured with the quartz microbalance deposition monitor.  The deposition monitor features a cover that is actuated with compressed air.  Both the outgassing rate and the deposition rate can be measured with the sample maintained at different temperatures, controlled by hermetic cartridge heaters mounted in the box platform.  Finally, within C2, the outgassing products from a sample can be collected on the glass disk mounted in the glass disk stage.

\begin{figure}
\includegraphics[width=3.5in]{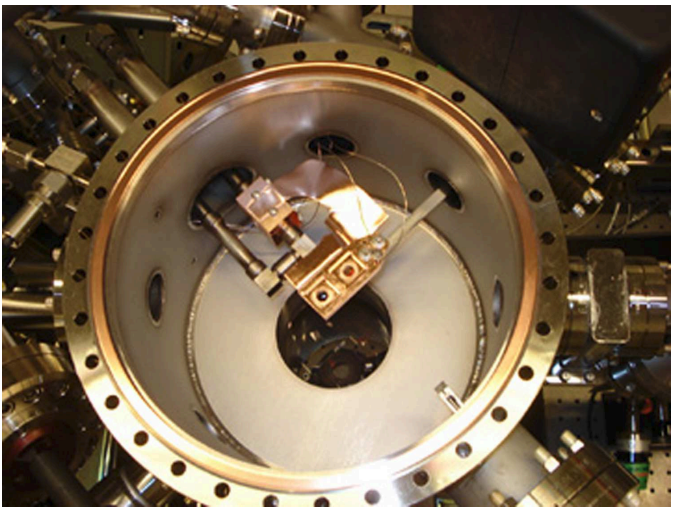}
\caption{Photograph of the interior of C3.  The double glass disk stage is seen on the end of an arm extending to the pneumatic actuator, with a thermal strap linking it to the cryogen loop.  The pincer of the wobble stick is visible at the lower right. }
\label{C3photo}
\end{figure}

Glass disks enter and exit through A3, and are moved via a stainless steel platform on the end of magnetic transport arm 3.  In C3 two disks are simultaneously placed in a double glass disk stage, maintained at -120$^{\circ}$C via a copper strap connected to a cryogen loop entering C3, and the light transmittance through both glass disks is measured in six different wavelength bands.  This done with both disks uncontaminated, and again subsequently with one disk contaminated by outgassing products in C2, to measure the light transmittance effects of contamination (see \S \ref{ltsec}).  Figure \ref{C3photo} shows a photograph of the interior of C3.  The double glass disk stage is seen mounted on the end of a pneumatic actuator which alternately places the contaminated and uncontaminated disks in the light beam, and the transmission through each is measured.

\section{Design details}

\subsection{Vacuum control and measurement}

\subsubsection{Pumping and performance} 

The main vessels are pumped primarily with a Varian Task-V301 combination turbomolecular and oilless roughing pump, with a capacity of 250 liters/sec for N$_{2}$ and a base pressure of 1x10$^{-9}$ torr.  All metal valves allow each vessel to be isolated from the turbo pump as needed.  C1 and C3 are connected to the main pump via 1" diameter vacuum lines, while C2 sees the pump through an in-line 6" diameter valve.  Pressures of 10$^{-9}$ torr are achieved in C2, while 10$^{-7}$ torr is achieved in C1 and C3 when empty.

The load lock vessels are pumped in stages, first with a scroll pumping system isolated from the main pumping, then through 1/4" pumping lines to the main pumping manifold, then via opening the load lock's gate valve to the adjacent main vessel, as shown schematically in Figure \ref{schematic}.  The oilless scroll pump is a Varian IDP3 with a capacity of 60 liters per minute and a base pressure of 0.5 torr.  Each main vessel also has an ion pump, which can supplement the turbo pump if necessary.  The ion pumps in C1 and C2 are Varian Valcon models, while the one for C3 is a Varian StarCell.

All flange seals are Con-Flat style with OFHC copper gaskets, with the exception of seals in the scroll pumping system, which are KF NW-16 with viton o-rings.  The gate valves between C1, C2, and C3 are all metal and manufactured by Varian, while the gate valves between A1 and C1, A3 and C3, and A2 and C2 are UHV metal valves with a viton bonnet seal, manufactured by MDC vacuum.  Valves to C1 and C3 on the main pumping manifold are all metal by Varian, while the valve to C2 is a 6" in-line crank operated all metal gate valve by VG Scientia.  VG Scientia is also the manufacturer of the wobble sticks with pincers.

A load lock undergoing pump-down reaches the base pressure of the scroll pump within several minutes, at which point it can be connected via narrow lines to the main pumping manifold and pressures of 10$^{-4}$ torr are achieved after approximately half an hour.  The main vessels can be pumped from atmospheric pressure to operating pressure in approximately one day

High vacuum pressures are read in the three main vessels with Varian inverted magnetron gauges.  Rough vacuum Varian ConvecTorr TC gauges read pressures in the three main vessels as well as in the main pumping manifold near the turbo pump.  All above gauges are read with a Varian Multigauge controller.  Pressures are monitored in the load locks with manual needle gauges.

\subsubsection{High vacuum techniques}

All vessels are stainless steel and underwent a 400$^{\circ}$C vacuum bakeout after fabrication.  The only materials used within the vessels, for all structures and fasteners, are OFHC copper and stainless steel, with the exception of the thermocouple wiring, which is necessarily chromel and alumel, and an indium gasket coupling to the cold loop in C3.  All seals and all structures and items within vessels are cleaned and prepared for high vacuum by the plating shop of the SLAC National Accelerator Laboratory.  Material samples for testing are cleaned beforehand in accordance with how the material would be prepared for the camera cryostat.  

Dividing the functions of the CMTC among three main vessels allows the lowest pressures and least potential contamination to be achieved in C2, where it is most beneficial for rate-of-rise and deposition measurements, and minimizes spurious contamination of the reference glass disk in C3.  The split system also keeps initial bake-out product contamination of structures in C2 and C3 to a minimum.    

Load locks, and main vessels when necessary, are brought from vacuum to atmospheric pressure with a dry nitrogen gas backfill, and a dry nitrogen purge is maintained for all vessels when at atmospheric pressure.  Nitrogen gas pressure is provided by the boil-off gas of a liquid nitrogen dewar, and distributed via a manifold which allows each vessel and load lock to be independently valved.  Vessels C2 and C3 feature 10 psig burst disks to prevent an overpressure situation while preserving vacuum integrity during operation.  A comprehensive overview of the theory and practice of high vacuum techniques is provided by, for example, Dushman and Lafferty\cite{SFVT}.

\subsection{Thermal control}

Temperatures of $\lesssim$-100$^{\circ}$C are needed for the glass disk stages and quartz microbalance deposition monitor, in order to replicate the collecting effects of the cold focal plane of the LSST camera.  The glass disk stages have thermal straps linked to cryogenic fluid loops that enter the vessels on tubes welded to 4 1/2" Con-Flat flanges with bayonette style feed-thrus and circulate cold fluid into C2 and C3.  The quartz microbalance has a separate fluid cooling loop that enters C2 on the same flange as the quartz microbalance readout and cover actuator.  The cryogenic fluid is circulated by a Polycold PFC1100HC refrigerator and copper refigerant line, with an insulted cryogenic valve box allowing the independent control of flow to each loop.

The refrigerant circulated from the Polycold unit is directed through a manifold with valves allowing flow to any combination of the C2 and C3 cryogen loops, and the quartz microbalance.  In C2, we use a group of eight 0.8 mm thick OFHC copper straps 2.5 cm wide and 64 cm long to link the glass disk stage to to the cryogen loop, and all connections are brazed joints.  To faciliate a brazed joint between the copper straps and the cryogen loop, the loop contains a copper section brazed between the stainless steel ones.  In C3, because of the need to acuate the glass disk stage and therefore for flexibility in the thermal link, we use one 0.8 mm thick OFHC copper strap of varying width which is curved into nearly a loop, with a total length of 18 cm.  The C3 cold strap is brazed to the glass disk stage at one end and brazed to a copper block at the other end; the copper block contains a half cylindrical cutout of the same cross section as the cryogen loop tube, and it is compressed against the tube along with another identical block and an Indium washer to reduce thermal resistance.  

Temperatures of -120$^{\circ}$C are achieved on both the C2 and C3 glass disk stages.  The fact that these cold stages, with mechanical and radiative couplings to room temperature and hotter surfaces, are only 15$^{\circ}$C warmer than the refrigerant liquid being circulated in the tubes indicates that the thermal conductivities achieved through the copper straps, braze joints, and Indium washers are quite high.  

The glass disk stage in C2 achieves this temperature even with the sample box stage heated to 90$^{\circ}$C just below.  We thermally isolate the two stages in C2 with ceramic washers on the four 4-40 stainless steel all thread standoffs that support the glass disk stage above the sample box stage.  

We have also used the C3 cold stage as a mechanism to cool candidate sorption pumping materials, which can then be moved to C2 to measure their pumping effectiveness at cold temperatures.  It should be noted that the rate of cryo-pumping to cold surfaces in the system is quite high, so any rate-of-rise measurements must be done in a vessel while there are no surfaces cold in that vessel in order to obtain an accurate result.

We heat the C1 heater block and C2 sample stage with hermetic DC heaters, two in each.  The C1 heaters are UHV disk heaters manufactured approximately 1.5 cm in diameter from Heatwave Labs, while the C2 heaters are 300 W cartridges aproximately 7 cm long and 1 cm in diameter manufactured by Chromolox.  Heater signals enter the vacuum on 2 3/4" Con-Flat electrical feed-thru flanges, and are powered with 5A DC power supplies.  Temperatures are measured with type-K thermocouples.  In addition to diagnostic thermocouples, there are thermocouples mounted on the heater block in C1, on the sample box stage in C2, on the C2 glass disk stage, on the C3 glass disk stage, and and on the quartz microbalance deposition monitor cooling loop.  Thermocouple signals pass out of the vacuum on thermocouple hermetic feedthrus with chromel and alumel pins mounted on 2 3/4" Con-Flat flanges and are processed by a Cole-Parmer 18200-40 USB thermocouple readout.  

\begin{figure}
\includegraphics[width=3.5in]{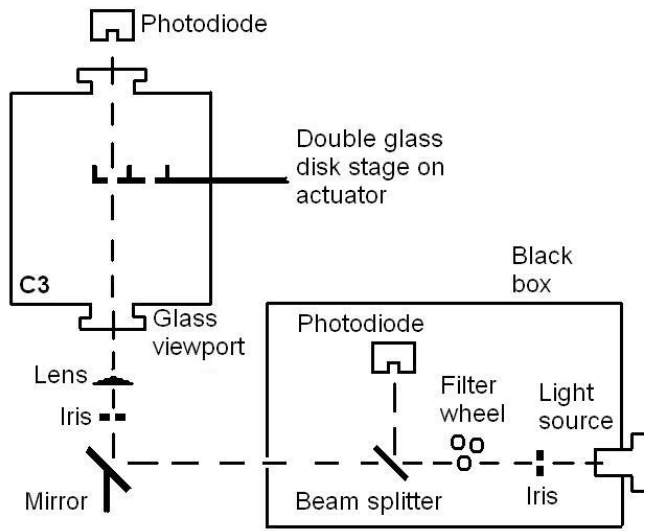}
\caption{Schematic of the optical system.  The light transmittance through the contaminated and uncontaminated glass disks is compared, as discussed in \S \ref{opt}. }
\label{optfig}
\end{figure}

\subsection{Optical system and peripheral devices} \label{opt}

For the optical system, we use a Newport full spectrum light source.  Within a black box mounted on an optical table, the light beam passes in sequence through an iris which serves to limit the width of the beam, a ThorLabs FW102 filter wheel with six filters approximately matched to the center frequencies of the six LSST camera filter bands, and a beam splitter.  The filters used in the CMTC are centered at 400, 500, 600, 750, 850, and 1000 nm and are the FB series by ThorLabs.  The filters at 370 and 1000 nm have widths of 10 nm while the others have widths of 40 nm.  Part of the light from the splitter is incident on a Newport detector photodiode, to be used as a calibration reference, which reduces the effects of intensity variations of the light source, while the rest is directed out of a small hole in the dark box and then by a mirror vertically through a lens and a glass viewport into C3.  

The C3 beam can pass through the contaminated and uncontaminated glass disks as described in \S \ref{pdv}, exits C3 through another glass viewport, and is then incident on a second Newport detector photodiode.  In this way, the transmittance in each band through the contaminated glass disk can be directly compared to the uncontaminated one, with the calibration reference diode used remove drifts in the inherent power of the light source.  Both detector diode powers are read with a Newport 2935-C power meter.  A schematic of the optical system is shown in Figure \ref{optfig}.

For glass disks, we use 15 mm diameter Edmund Optics BOROFLAT Borosilicate windows which are 1.75 mm thick.  The disks are held in a specially designed two piece OFHC copper holder which pinches the glass disk from both sides, assuring a thermal link between the glass and the copper.  The flat bottom of the holder fits into cutouts in the glass disk stages in C2 and C3, in this way providing for thermal conductance between the glass disk and the cold stages.

The power output by the diodes is read by a Newport 2935-C power meter.  The double glass disk stage in C3 is precisely moved to alternately allow each glass disk to be in the same position in the light beam with an MDC ACBM-275 pneumatic actuator.  We produce the actuator switching logic signal from computer software with a National Instruments SCB-68 box, and the logic is run through a Continental Industries relay.  The relay voltage output is the control signal for a Humphrey 410-39-70 solenoid control valve, which controls the flow of compressed air to the actuator itself.

The quartz microbalance and interface box are the Inficon XTM/2 system.  We interface the Multigauge controller, quartz microbalance, RGA, pneumatic actuator, and diode power devices to two PCs through the LabView environment.  We used compressed air to provide pressure actuate the quartz microbalance lid and the linear actuator which moves the glass disk stage in C3.

\section{Evaluation of experimental data}

\subsection{\label{rorsec}Rate-of-Rise technique for outgassing rates}

\begin{figure}
\includegraphics[width=3.5in]{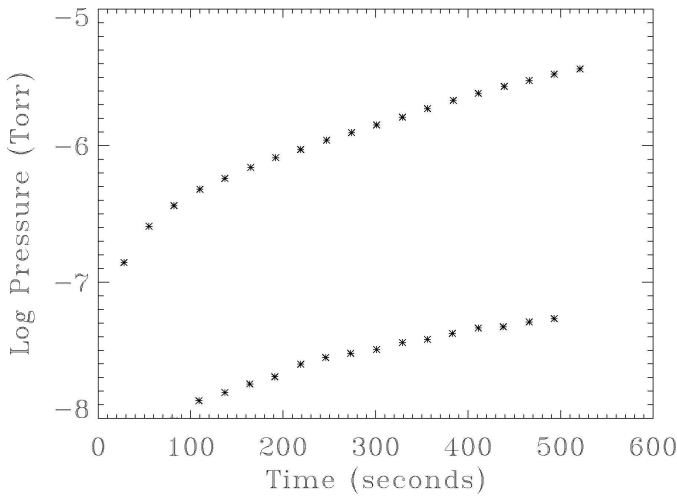}
\caption{Partial pressure vs. time of molecular weight 18 (water) for a sample of proprietary epoxy at 90$^{\circ}$C, as measured in the CMTC.  The upper data points are with the sample, and the lower data points are the background level.  The slope of a line fit to the points (neglecting the first several) gives the rate-of-rise in torr/sec.  At this temperature, even for a very dry sample such as this, the rate of rise including the sample is nearly two orders of magnitude larger than just the background.  }
\label{rorfig}
\end{figure}

Gregory\cite{ror} has demonstrated the validity of the rate-of-rise technique for measuring outgassing rates.  If a vessel is isolated from pumping, then the outgassing rate Q of a given molecular species i is given by 

\begin{equation}
Q_{i}  = V\frac{dp_{i}}{dt}\,,
\label{roreqn}
\end{equation}

where dp$_{i}$/dt is the rate of change with time of the partial pressure of molecular species i.  With the partial pressures of each molecular weight read by the RGA, we can achieve outgassing rate measurements by fitting a line to partial pressure versus time for each molecular weight for both a background level and a level including the sample, and subtracting the former slope from the latter.  Figure \ref{rorfig} shows the rate-of-rise data points for molecular weight 18 (water) for a sample of a proprietary epoxy to be used in hermetic feed-thrus in LSST at 90$^{\circ}$C as measured in the CMTC.  When multiplied by the volume of the chamber and divided by the surface area of the material, the resulting outgassing rate per unit surface area in torr-liters/sec/cm$^{2}$ is the standard form of rate-of-rise reported for materials.  It can be used to determine the outgassing rate of any quantity of the same material in the camera cryostat, assuming a direct scaling with surface area.

\begin{figure}
\includegraphics[width=3.5in]{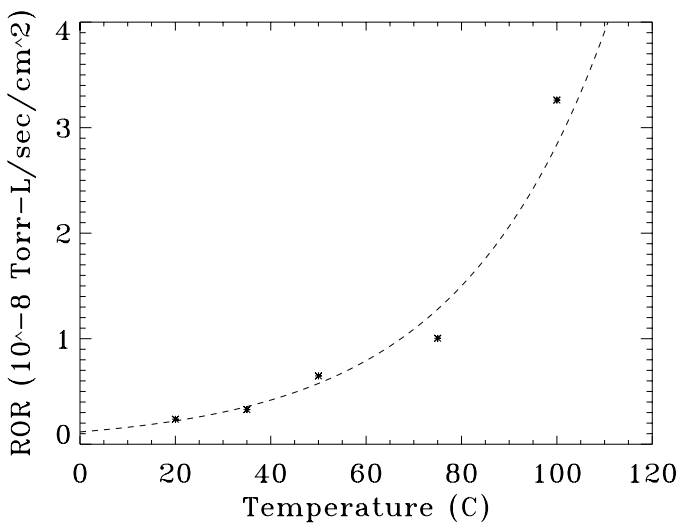}
\caption{Outgassing rate of water versus temperature for a sample of FR4 circuit board.  The dashed line is the exponential fit of the form given in equation \ref{roreqn2}.  }
\label{rorfig2}
\end{figure}

We typically measure outgassing rates with the material sample at a range of temperatures.  This allows a first order determination if any potentially reactive products are present in the material sample.  For samples with significant outgassing, one can fit an exponential function of the form 

\begin{equation}
R_{i}(T)  = A_{i} e^{b_{i} T}\,,
\label{roreqn2}
\end{equation}

where R$_{i}$ is the outgassing rate of the ith chemical species and T is the temperature in Kelvins.  This functional form allows the outgassing rate to be extrapolated to any temperature, in particular to the temperature a given component will be during normal LSST camera operation.  Figure \ref{rorfig2} shows the outgassing rate of water for a sample of FR4 circuit board measured at different temperatures.  We can then add the outgassing rates of chemical species that will condense on the focal plane at its operating temperature and those that won't, for an estimate of the amount of deposition on the focal plane resulting from a given material, and an estimate for the overall gas load, respectively.

There is data in the literature concerning rate-of-rise measurements for substances.  For example, Wong\cite{rorlit} maintains a database of reported rate-of-rise measurements.  As is evident there, reported values can vary over orders of magnitude for the same bulk substance at the same temperature, indicating the importance of the effects of both microscopic surface finish properties and material preparation techniques (eg. cleaning, baking, handling) on the rates-of-rise from a sample.  Our present effort is to measure materials with the same surface finish and preparation as they will undergo for the LSST camera.  Given the dispersion in the literature data, and the importance of varying finish and preparation, we find it difficult to estimate uncertainties in CMTC rate-of-rise measurements via comparison with literature data.  For estimates of uncertainty in rate-of-rise measurements then, we believe that repeated measurements with identical samples are the appropriate technique.  Based on our experience thus far, the uncertainty in the rate-of-rise determined in this way varies with substance, however, we believe that a factor of two in torr L$^{-1}$ sec$^{-1}$ cm$^{-2}$ is an appropriate conservative bulk figure of merit for the system.

\subsection{Deposition rates}

\begin{figure}
\includegraphics[width=3.5in]{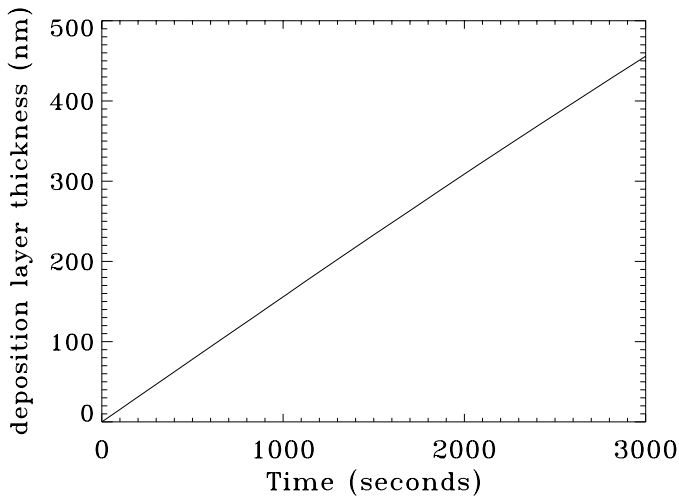}
\caption{Thickness of the deposition layer of water on the quartz microbalance deposition monitor versus time for a sample of getter material which is water saturated.  This sample intentionally has a high rate of outgassing.  The deposition monitor is maintained at a constant temperature near -100$^{\circ}$C, and is located approximately 1" below the outgassing sample.  Given the surface area of the quartz crystal, and that the deposition product in this case is water, the mass deposited is read directly by the deposition monitor and converted to an average deposition layer thickness.  }
\label{qbfig}
\end{figure}

We use the quartz microbalance deposition monitor to estimate the rate at which products outgassing from a sample condense on a cold surface.  The oscillation frequency of the quartz crystal at a given temperature is solely a function of the mass, and the frequency is read and converted to a mass deposited with the Inficon XTM/2 interface.  We maintain the quartz microbalance at $\sim$ -100$^{\circ}$C to simulate the deposition onto the cold camera focal plane.  The temperature of the quartz crystal must be  constant to within a fraction of a degree to avoid the oscillation frequency dependence on changing temperature.  With an estimate of the mass deposition rate and the surface area of the sample, this can be scaled up to an estimated mass deposited on the focal plane in a given time due to a given quantity of material in the camera.  With knowledge of what the outgassing species are, the density of the material and surface area of the focal plane allows an estimate of the deposition layer thickness.  We find in general that only samples with outgassing rates of condensable species much higher than that expected for materials in the LSST camera show appreciable deposition, given the small sample sizes allowed.  Figure \ref{qbfig} shows deposition from a sample of a getter material which is water saturated.

\begin{figure}
\includegraphics[width=3.5in]{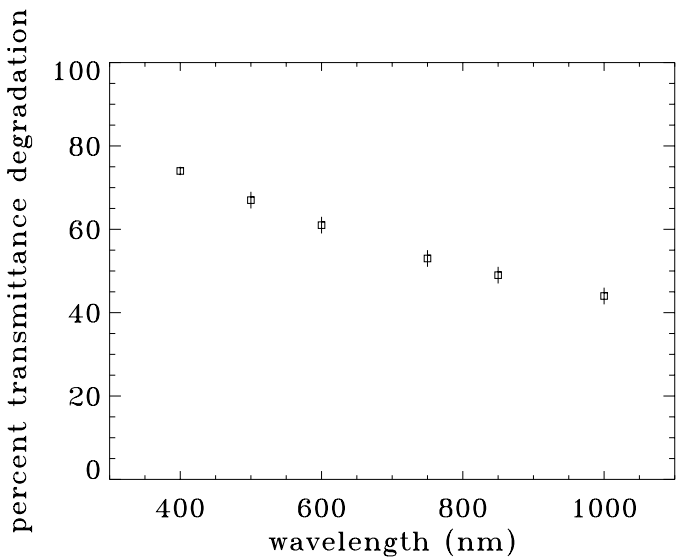}
\caption{Percent degradation in the light transmittance through a glass disk maintained at -120$^{\circ}$C and exposed to the outgassing from a sample of water saturated zeolite pellets for 24 hours.  This sample intentionally has a high rate of outgassing, to demonstrate a large effect.  The reported errors are statistical, and are at the level of 1\% of degradation for the band centered at 400 nm and 2\% for the other wavelength bands.  We reduce systematic error by the design of the optical system and by comparing transmission through the disk before and after contamination to that through an uncontaminated reference disk, as discussed in \S \ref{ltsec}.  From standard diffraction and absorption theory, it is expected that effects should be enhanced at shorter wavelengths. }
\label{opticalfig}
\end{figure}

\subsection{Light Transmittance}\label{ltsec}

In C3, we measure the light transmittance through two glass disks on a double cold stage moved by a pneumatic actuator, in six optical bands, as discussed in \S \ref{pdv} and \S \ref{opt}.  First the transmittance through both disks is measured with both disks uncontaminated, and then it is measured after one disk has been exposed to outgassing products in C2, as in \S \ref{setup}.  In a given measurement sequence, for each of six filters the transmittance through each disk is sampled 20 times, and the sequence is generally repeated five times.

All raw transmission values through a given disk are expressed a ratio, as the power incident on the diode above C3 is divided by the power on a reference diode.  We form a further ratio of the transmissions through the two disks before the contamination of one disk and then after, and divide the later ratio by the former.  This gives a value for the fractional degradation in transmittance through the contaminated disk, with common mode changes to the system normalized out.  We determine the statistical uncertainty in transmittance degradation by repeating each raw transmission measurement several times and taking the standard deviation, then using standard error propagation to achieve a value for the uncertainty in the calculated transmittance degradation.

Figure \ref{opticalfig} shows the percentage transmittance degradation through a disk exposed to the outgassing products of a sample of water saturated getter material for 24 hours.  This sample intentionally has a high rate of water outgassing, so the degradation is high.  From standard diffraction and absorption theory, it is expected that transmittance degradation effects should be enhanced at shorter wavelengths for this level of deposition.

\begin{acknowledgments}
We thank Howard Rogers for fabrication expertise.  We also thank Leo Manger for his help, and Ali Farvid and colleagues in the SLAC plating shop.  
\end{acknowledgments}

\end{document}